# CRITICAL INFRASTRUCTURE CYBERSECURITY CHALLENGES: IOT IN PERSPECTIVE

[1]Akwetey Henry Matey, [2]Paul Danquah,
[1]Godfred Yaw Koi-Akrofi and [1]Isaac Asampana

[1]Departments of I.T. Studies, University of Professional Studies Accra
[2]Department of I.T., Heritage Christian University

## ABSTRACT

*A technology platform that is gradually bridging the gap between object visibility and remote accessibility is the Internet of Things (IoT). Rapid deployment of this application can significantly transform the health, housing, and power (distribution and generation) sectors, etc. It has considerably changed the power sector regarding operations, services optimization, power distribution, asset management and aided in engaging customers to reduce energy consumption. Despite its societal opportunities and the benefits it presents, the power generation sector is bedeviled with many security challenges on the critical infrastructure. This review discusses the security challenges posed by IoT in power generation and critical infrastructure. To achieve this, the authors present the various IoT applications, particularly on the grid infrastructure, from an empirical literature perspective. The authors concluded by discussing how the various entities in the sector can overcome these security challenges to ensure an exemplary future IoT implementation on the power critical infrastructure value chain.*

## 1. INTRODUCTION

The power generation and distribution sector has seen tremendous growth infrastructure-wise due to technological advancement in IoT. The inculcation of IoT in the critical infrastructure has enabled applications such as load scheduling, routing, and exchange of information through telematics and self-healing processes to extend the life of the grid infrastructure. IoT's advanced remote sensing capabilities provide more accurate views on generation capacity and visualization of summary information from these sensing devices [1-3]. It encourages decentralized systems in energy generation from a renewable energy perspective such as wind, solar, etc. [4, 5]. It also supports advanced insight into the utilization of electricity infrastructure [1]. The real-time nature of the critical infrastructure espouses a more sustainable distributed power generation in self-directed systems to react dynamically to changes in power demand and distribution, respectively.

In realizing the full potential of these benefits, it is equally important to identify the security implications of IoT platforms in the critical infrastructure.

Despite its essential role in the power generation and distribution sector, there are resource constraints and dynamism of its network infrastructure. Leaving participants to be faced with challenges [6, 7] indicated that IoT integration into the energy market bridges the gap between personnel and the energy infrastructure, of which cybersecurity challenges can occur. As confirmed by [8], the emerging concept of cyber-physical (CPSs) and IoT introduced in the energy market will cause the system to be vulnerable to cyber-attack. Even Interconnecting different devices developed with other protocols and standards in a single IoT platform can pose a significant challenge despite its efficient implementation and utilization on critical infrastructure [9]. Figure 1 below illustrates an IoT electric Power Network.

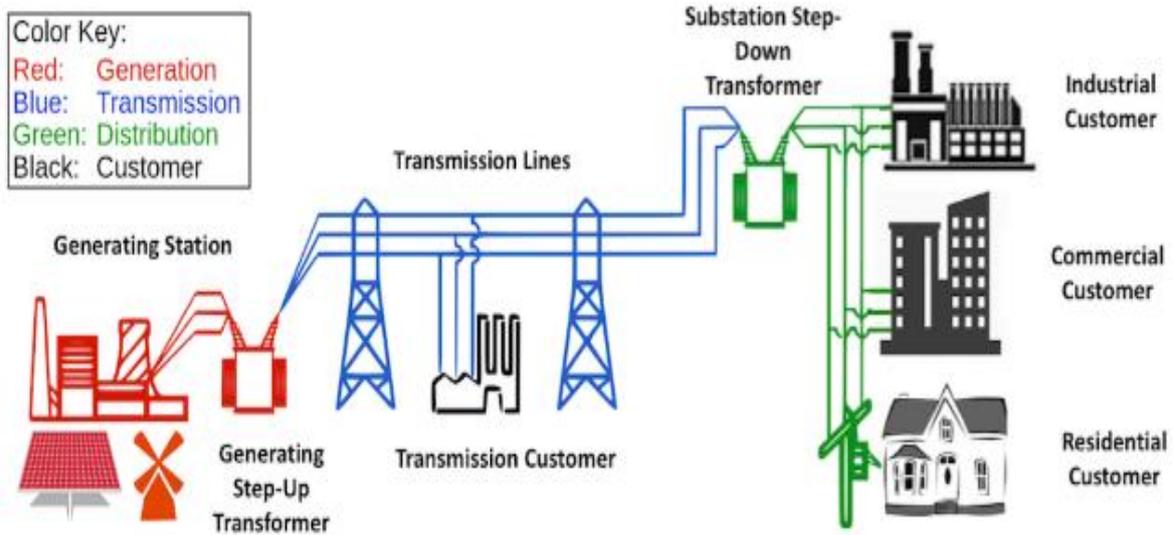

Figure 1. Electric power networks

Sources: In [10]

From the narratives above, the current study seeks to review the implementation challenges from existing literature on IoT application on the critical infrastructure of the power sector. Hence the objectives of the study are to:

1. Discuss the current Energy Infrastructures Systems.
2. Identify IoT security challenges in line with the critical infrastructure
3. Discuss the various IoT security challenges and how to overcome these challenges.

IoT application on the critical infrastructure has called for an explosion of cybersecurity challenges on the grid. Therefore it has become necessary for authors to review the current IoT-based literature on critical infrastructure about cybersecurity challenges to suggest possible resolutions to mitigate such security challenges from a literature perspective. The authors present a discussion on IoT application deployment in line with the grid. Hence the rest of this paper is organized as follows:

  i. Highlights on IoT Energy Infrastructure
 ii. Insight on the grid perspective in line with IoT-based system architecture
iii. Discussion on IoT security challenges and how to mitigate these security challenges; and finally
 iv. Discuss the conclusion and recommendation for the study.

Without suitable security, the grid cannot play a significant role in the power management system [11].

## 2. METHOD

In conducting this systematic review, we adopted Bendermacher et al. [12] methodological approach to (i) develop a search strategy for using several databases, (ii) define exclusion and inclusion criteria for publications – assess for eligibility, (iii) define review and coding scheme, (iv) analyze and synthesize data and (v) develop write-up. To ensures transparency and rigor regarding the publication selection and analysis process. The delimited database search was keywords, abstract, and title. Given the literature volume, we adopted this strategy to reduce the number of publications to review while enhancing the precision of information search [13]. The goal was to identify relevant publications that explicitly discussed the concept of IoT cybersecurity challenges on energy infrastructure as a central thesis [14].

We reduced Research publications numbers by specifying the criteria for inclusion and exclusion. The inclusion criteria were publications on conference papers and articles (conceptual and empirical studies) in peer-reviewed journals written in English.

Furthermore, we exclude book chapters, reports, policy documents, newspapers, and magazine reviews from our sample.

The authors restricted the publications to 2017 to 2021, including vital definitions, concepts, and relevant information relating to the subject matter. The pre-selection strategy used ensured that relevant publications made significant contributions to the phenomenon under investigation included in the systematic literature review process. Also, we examined abstracts, keywords, introduction, and conclusion of each article as a means of reducing selection errors. The eligibility assessment involved manually screening each publication to enhance the rigor, accuracy, and reliability of the publication selection process.

The researchers used only secondary data, which refers to data already collected for some other purpose [15]. Secondary information was helpful for this study's purposely for analyzing the literature on cybersecurity. To identify relevant publications to the concept investigated by exploring twelve (12) databases. These include Springer, Association Information System (AIS) library, SAGE Journals, Scopus, IEEE Xplore, Association of Computing Machinery (ACM), Google Scholar, ResearchGate, Academia.edu, Emerald (database), Elsevier (database), and Pro-quest. The researchers used search string to search for publications in the various databases.

Researchers retrieved 707 publications from database searches, 350 of which were chosen based on title analysis. A count of 152 papers was deemed irrelevant after analyzing the abstracts of the publications. Also, 65 articles were taken out due to duplication, leaving 133 publications. Furthermore, 25 articles were excluded based on the criteria for inclusion and exclusion because they did not match the research aim. Then, the authors took out 28 publications after analyzing the text of the complete publications because the central focus of such publications was not on the concept of project management competencies, leaving a total of 80 sample sizes for detailed review and synthesis. It is worth noting that even though the authors worked with 80 articles, not all of them were so useful in the analysis.

## 3. RELATED WORKS

### 3.1. IoT Energy Infrastructure

IoT is an agent for change in transforming industries in the world. Depending on the problem domain to be addressed, its application usually focuses on the sub-domain of IoT implementation of the power sector. [16] Argues the need to unify methodology on the industrial standard to simplify the IoT architecture in the power sector. [17] also indicated the need for common technical ground to enhance interoperability. Meanwhile, [18] reasoned that future objects and devices would be connected and managed with the help of communication networks which are cloud-based servers. The heterogeneous nature of most architecture platforms ( [17], [19]–[21] ) gives indications of the variations in domain requirement and performance expectations of IoT critical infrastructures. Earlier research on critical infrastructures proposed a concept of abstraction ranging from the Industrial Internet Reference Architecture (IIRA) and Internet of Things Architecture [16], without much focus on security. [22] Proposed the open systems IoT reference Model (OSiRM), which is different from the one developed by International Telecommunications Union (ITU). Hence the need for a highly efficient communication architecture is imperative [23].

### 3.1.1. IoT smart grid perspective

IoT plays a significant role in critical infrastructure by reducing frequent visitation to plant and reducing human interventions in monitoring systems locations from the control center.

It also monitors electricity generation of different kinds of power plants, measuring various parameters, interoperability between other networks, monitoring to discover fault issues, eliminating them, collecting data, measuring abnormality, monitoring electricity quality, etc. [24-25]. Fundamentally, the IoT role encompasses monitoring the environment through actuators [26]. The grid architecture represents an advanced metering infrastructure with an enabling two-way communication. Figure 2 below gives a representation of IoT based distribution network.

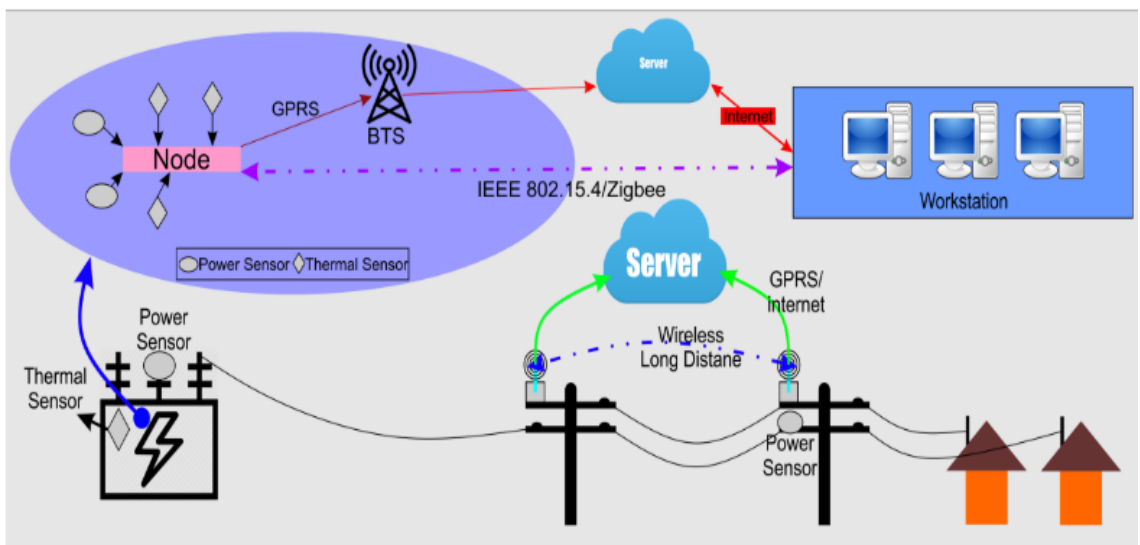

Figure 2. IoT Based Distribution network

**Sources:** [27]

An essential requirement of the critical infrastructure is to provide clean and reliable power to utilities, of which the following are key components: Cybersecurity, availability of adequate storage technologies, Data Management, communication networks (GSM, GPRS, ZigBee, PLCC, and broadband), System stability synchronization of distributed sources enabling bidirectional power flow, power, and electric vehicle management [28]. Not all IoT architectural vulnerabilities are addressed with its rising ubiquity due to resource restrictions [29].

International Journal of Network Security & Its Applications (IJNSA) Vol.13, No.4, July 2021

Power generators, massively use years back, Supervisory control and data acquisition systems (SCADA). Today, the generation and distribution of power depend on IoT Applications for connecting key areas in tuning operational activities of the power plants and balancing production cost of maintenance of equipment life span [30].

Therefore Smart Sensor Technology (SST) becomes an integral part of the intelligent grid, informing the control system about specific parameters and the happenings on a physical object on the grid [27]. Others for power quality monitoring, line monitoring, conductor temperature sensing, magnetic field sensing, strain sensing, accelerometer sensing for vibration, etc. [31]. These sensors also provide a technical solution to achieve a high level of accuracy on system quality and reliability[32]. Meanwhile, the lack of technical barriers is making it challenging to determine a clear policy direction due to the diverse characteristics of each sensor [33]. However, IoT promotes a better understanding of the business process, inspires business analytics of grid activities, and reduces unavailability in minutes of interruption [34].

### 3.2. IoT Application and Security Challenges

Cyber-attacks are artificial manipulation on the power grid to redirect power flow to an unassigned destination [35]. Attackers may corrupt or block information to either increase or decrease the value of power [36]. An instance of a cyber-attack on load frequency control (LFC), causing it to be unstable [37]. A recent study on industrial control systems by [38] reviewed Internet threats being distributed via advertising banners and phishing websites and in various intrusive hardware as detailed in table 1 below:

Table 1. Percentage of threats blocked on ICS

| Type of treat       | Percentage of ICS (%) |
|---------------------|-----------------------|
| JavaScript Trojans  | 3.1                   |
| JavaScript Miners   | 1.2                   |
| Spyware             | 0.7                   |
| Malicious document  | 0.5                   |
| Web Exploits        | 0.1                   |

Sources:[38]

With the increasing development of cybersecurity treat in the power sector, there is the need to pay particular attention when integrating wireless sensor networks (WSN). With IoT, services mechanisms, users acceptance, and data privacy management [39-40], designers of systems have their concerns in designing a particular architecture.

Such alarms call for a unified architecture for IoT-based energy systems, which researchers need to focus on in the area [7]. The main challenge of enabling IoT in energy systems is to map every object into one unique virtual thing. Meanwhile, because the IoT sensors lack critical features such as situational intelligence, efficient power management, and cybersecurity features, there is the need for incorporating such elements into future IoT sensors to enhance their functionalities [41]. [16] Also indicated the importance of Low-power wireless network security, which is an essential factor to consider when implementing IoT. According to [37], cybersecurity features to control systems against unauthorized access and mitigation in an isolated power station is an important area to be considered by the research community. [42] outline the critical threats on the hybrid power plant, namely, Distributed Denial of Service (DDoS), false data injection (FDI), Compromised Key, Man-In-Middle, Replay Crash Override, Packet Drop attack, Jamming attack,

45

Electronic copy available at: https://ssrn.com/abstract=3910922

and Stealthy Deception. A recent cyberattack report by [43] was an attack breached of the industrial control systems (ICS) the case of U.S. and other countries 2016 and 2017 using "Dragonfly". Secondly, an attack on a small cloud service impacted the natural gas, oil, and electric power sectors in the United States. Finally, there was the NotPetya attack which halted various operations globally across multiple sectors in 2017. Potential attack surfaces in the power grid are Data concentrator(D.C.), SCADA, Control System, State estimator, Communication channel, Power market**,** Remote terminal unit(RTU), Phasor measurement unit(PMU), Programmable logic controller(PLC), and Advanced meter infrastructure (AMI) [35]. Table 2 below gives a detailed account of some cyber incidents and target categories on the grid.

Table 2. cyber-Incidents and Target/surfaces

| Author/Authors | Cyber Incidents | Target/ Surface Categories |
| --- | --- | --- |
| [44] | Spreading of Malware in grids | Smart grid environments |
| [45] | Malicious data attacks on measurements | Smart grid |
| [46] | Non-stationary adversarial cost variation constraint | Smart grid |
| [47] | The lack of chronological dynamic attack models to minimize system damage remained unexplored. | Smart grid |
| [48][49] | False data injection (FDI) attacks | Phasor measurement units (PMUs). Real-time phasor measurements. |
| [50][51] | Unreliable data transmission | Phasor measurement unit (μ-PMU) Phasor measurement units (PMUs). |
| [52] | False data injection (FDI) attacks | IEEE 39-bus and the actual provincial backbone network |
| [53] | Stealthy false data injection (FDI) | State Estimation |
| [54] | False Data Injection (FDI) | State Estimation Component |
| [55] | False Data Injection Attack (FDIA) | State Estimation Component |
| [56] | False data injection (FDI) into meter measurements | Security constraint and economic dispatch (SCED) |
| [57] | False Data Injection Attack (FDIA) | IEEE 14-bus and 118-bus systems. |
| [58] | Injection of false measurements rendering it to cause overloads. (FDI) | Traditional data detection state estimation |
| [37] | Unstable operation by unauthorized access | Load frequency control (LFC) |
| [59] | There is a Price manipulation attack and Denial of Service (DoS) attack. | Issue of Granular price changes, monitor, and another node in the network. |
| [60] | Denial of service (DoS) attacks and malicious injection code | Smart grid |
| [61] | Denial-of-service attacks (DoS) | Grid Electrical Vehicle Batteries (GEVB) |
| [62] | False data injection (FDI) and denial of services (DOS) attack to falsify the sensor measurements | Automatic generation control (AGC) |
| [63] | Misrepresent the sample value (S.V.) messages | IEC 61850 protocol |
| [34] | The inadequate energy efficiency of | IoT Devices |

|  | IoT platforms |  |
| --- | --- | --- |
| [36] | Blockage or corrupt information to increase or decrease the value of power | Metering that sends dummy value to the control room |
| [64] | Unreliable in storing excessive generated energy. | Battery Energy Storage Systems (BESSs) |
| [65] | Eavesdropping, Denial of service attack, forgery of instructions, tampering of measurement data, hijacking, and interference | Power generation acquisition terminal of a new energy plant (PGATNEP). |
| [66] | Coordinated Malware maliciously employed to turn on or off high-wattage appliances synchronously. | Simulate Coordinated Load-Changing Attacks (CLCA) on IoT platforms |
| [67] | Malware Attack | Industrial Control System (ICS/SCADA protocols. |
| [34][38] | Access to corporate and internet email services | Attack on Industrial Control System (ICS) and computers in power organizations. |
| [68] | Man-in-the-middle attack | Generic Object- Oriented Substation Event (GOOSE) protocol of IEC 61850 protocol |
| [69] | Lack reliable evaluation framework for multi-state power systems to reserve capacity. | Demand-side resources (**DSRs)** are responsible for providing reserve capacity to enhance the power system's reliability. |
| [35] | Adversarial attacks and cyber intrusions | Automatic Generation Control (AGC), State Estimation (S.E.), and Energy Management Systems (EMS). |
| [70] | Coordinated attacks on power flow | IEEE 14-bus distribution system and the IEEE RTS79 system |
| [71] | Corrupted data on components based on fault injection | IEEE 14-, 30-, and 57-bus |
| [72] | State attack and sensor attack | Cyber-physical power systems (CPPS) |
| [73] | Distributed Denial of Service attack | |
| [74] | Dynamic load altering attacks (D-LAA) | |
| [75] | Phishing attacks, credential theft, distributed denial-of-service (DDoS) attacks, kill Disk attacks, and unauthorized Remote Terminal access. | Substation Automation Unit (SAU), state estimation, Volt VAR control, and FLISR |

## 4. ANALYSIS AND DISCUSSION

### 4.1. Cybersecurity Challenges in the Power Generation Sector

Based on our empirical review, the principal cybersecurity challenges we identify were base on the years 2017 to 2021. The authors recognized that security is evolving, focusing on the selected years to keep up with the current cyber-attack incidents on the grid, as indicated in table 2. Our review reveals various attack incidents on the critical infrastructure regarding introducing different IoT on the grid infrastructure.

## 4.2. Cyber-Attack Challenges and Countermeasures

In table 3, the authors identified the following cyber incidents: false data injection (FDI) attacks, denial-of-service (DoS) attacks, distributed denial-of-service (DDoS) attacks, man-in-the-middle attacks, malware attacks, state estimation attacks, code injection, dynamic load altering attacks (D-LAA), unauthorized access to systems, eavesdropping, phishing attacks, and killing disk attacks.

Table 3. Type of Attack

| Author/Authors | Type of Attack |
| --- | --- |
| [36], [48], [49], [52-58], [62] | False data injection (FDI) attacks |
| [59], [60], [62], [75][65] | Denial-of-service attacks (DoS) |
| [73], [75] | Distributed Denial of Service attack |
| [68] | Man-in-the-middle attack |
| [44], [66], [67] | Malware Attack |
| [36], [60], [71] | Code injection |
| [73] | Dynamic load altering attacks (D-LAA) |
| [34], [37], [38], [75] | unauthorized access to systems |
| [45], [62] | State Estimation attack |
| [65] | Eavesdropping, Phishing attacks & kill Disk attacks. |
| [59], [63] | Misrepresentation of values and Price manipulation attack |
| [66], [70] | Coordinated attacks |

Below, we discuss these major cyber incidents identified in Table 3 and their related Countermeasures.

## 4.3. False Data Injection (FDI) attacks

A novel distributed host-based collaborative detection method was proposed by [48] in an event where the phasor measurement units (PMUs) are compromised. Simulation for detecting and isolating cyber-attack using real-time synchrophasor measurements was also proposed by [49]. Considering the attack on a critical node, [80] also proposed a model to investigate vulnerabilities subject to target sequential attack in node importance. On state estimation attack, [81] indicated the need to apply multiple metrics to monitor abnormal load deviations and developed a graphical detection technology that uses Graph Network (G.N.) to detect tampered measurements. To resolve such incidents, sometimes, a data-driven machine learning-based scheme can be used [53]. A Framework for measuring gross error [54]analysis was also deployed by [57] for processing and analyzing FDIs. [62] propose a mitigation platform (CDMP) for detecting cyber-attack, which uses forecasted data to investigate multiple generations and distribution companies under bilateral trading. This CDMP is accomplished by combining historical data from the IEEE 118-bus system with the dynamic analytical framework (DAF) analysis of potential cyber-attacks intra-interval operational security impacts.

## 4.4. Denial-of-Service Attacks (DoS)

For a Denial-of-service attack in the phase of price manipulation,[59]proposed an Intrusion Detection System (IDS) architecture in addition to a Cumulative Sum (CUSUM) algorithm to detect granular price changes, monitor and detect mischievous nodes. [60] Also proposed the act of detection-based defense and protection-based defense in the form of categorization of the International Journal of Network Security & Its Applications (IJNSA) Vol.13, No.4, July 2021attack. To detect DoS, CDMP was proposed [62]. For Target on-grid electrical vehicle batteries, [46] proposed a multimodal vibration countermeasure to DoS. [73] Indicated Fog computing capabilities, which serve as a layer between insight and cloud, for performance enhancement and to execute delegated tasks on behalf of the cloud. To ensure the availability of core Distribution Grid Automation (DGA), an automatically distributed approach in line with Blockchain and Smart Contract is proposed by [75] in the phase of distributed denial-of-service (DDoS).

### 4.5. Man-in-the-Middle Attack

In the event where the attack is Malware, [44] calls for an all-inclusive, generic model on cyber-attack life-cycles to address specific grid environments. Meanwhile, when such an attack is said to be coordinated, [66] believes that Coordinated Load-Changing Attacks (CLCA) are simulated on the grid with various power plants. Where this Plant is under normal or under attack situations. [67] Also suggested that when such an attack is target towards Industrial Control System (ICS), system monitoring monitors ICS/SCADA protocols about I.P. flows extended to application layer data obtained from ICS packet headers.

In table 4, we provide details of other specific cyber-attack challenges on the grid network.

Table 4. Specific cyber-attack challenges on the grid

| Author(s) | Nature of attack | Countermeasures |
|---|---|---|
| [34] | The energy efficiency of IoT platforms | The energy efficiency of the devices is gain through knowledge extraction from data collected in the early stages. In this manner, a large amount of data is the move to avoid latency. |
| [45] | malicious data attacks on measurements | The method required to detect, identify and correct malicious data attacks in intelligent grids |
| [37] | Cyber-attack on the load frequency control (LFC) | The need to ensure compliance of cybersecurity on the control system against unauthorized access. |
| [63] | Cyber-attacks on misrepresentation of sample value (S.V.) messages of the IEC 61850 protocol, | Use of an algorithm based on continuous monitoring of electric power system (EPS) parameters to detect cyber-attacks or to identify the most vulnerable elements on the electric network, The need to redefine personnel operational behavioural strategy |
| [36] | Cyber-attack obstructs or corrupt information to either increase or decreases the value of power. | The concept of dummy value is applied. Where meters in the smart grid send dummy values to the control room and the actual value that is most likely the essential reading, such an approach is safe against false data injection attacks or stealth attacks. |
| [64] | Cyber-attack against Battery Energy Storage Systems (BESSs) | Use smart contracts to control distribution and improve the security of BESSs by utilizing a replicated execution and state consensus features and |

49

| | | | cryptographic techniques. |
|---|---|---|---|
| [65] | | Eavesdropping, Denial of service attack, forgery of control instructions, tampering of measurement data, hijacking, and interference on the Power generation acquisition terminal of a new energy plant (PGATNEP). | Use of threat monitoring of PGATNEP to effectively improve the information transmission security and protection capability of power generation acquisition terminals.<br>Additionally to the above, the need for the adoption of business isolation for general high-risk nodes. |
| [34][38] | | Access to corporate and internet email services of electrical engineering workstations | Avoid using external Electronic mail services on industrial control systems (ICS) in power generation organizations.<br>Corporate email service should be restricted on ICS computers in power and energy organizations and only allowed in cases where necessary.<br>Secondly, the need to<br>• Manage the rights of user and service accounts, restrict access, Audit the use of privileged accounts, and regularly review access rights |
| [68] | | Exploitation of cyber vulnerabilities on Generic Object- Oriented Substation Event (GOOSE) protocol of IEC 61850 protocol with a man-in-the-middle attack. | By ensuring the authenticity and integrity of message using authentication codes at the end of every Generic Object-Oriented Substation Event (GOOSE) message, as standardized by IEC 62351- 6. |
| [69] | | Cyber-attack on Demand-side resources (DSRs) being responsible for providing reserve capacity and enhancing the reliability of power systems. | Proposed an innovative operating Framework for evaluating multi-state power systems for DSRs, particularly when considering cyber system failure.<br>When conducting optimal power flow on an innumerable states system, it is crucial to use a reliability index based on load restriction. |
| [35] | | Cyber-attacks are target towards key power system operational functions such as automatic generation control (AGC), state estimation (S.E.), and energy management systems (EMS). | Off-the-shelf cyber intrusion detection techniques can strengthen ICS and protect the intelligent power grids against malicious cyber-attacks. |
| [50] | | Cyber-attacks on the Micro phasor measurement unit (-PMU) rely on communication networks for data transmission. | To reduce the likelihood of an attack, an intelligent island detection approach on u-PMU is used. |

| [70] | Attack on the IEEE 14-bus distribution system as well as the IEEE RTS79 system. | Use a process based on Nested Column-and-Constraint Generation (NCCG) algorithm with the duality-based method to derive optimal solutions. |
| --- | --- | --- |
| [46] | Non-stationary adversarial cost variation constraint on the grid to have optimal intelligent grid protection against cyber-attacks in a reasonably practical scenario. | Thompson–Hedge algorithm can resolve the problem; superior performance of the proposed algorithm in terms of the convergence rate of regret function. |
| [71] | Fault injection on IEEE 14-, 30-, and 57-bus | Simulation-based on fault injection to evaluate survivability from corrupted data. |
| [72] | State attack and sensor attack in cyber-physical power systems (CPPS) | Application of detection logic to detect state attacks and sensor attack detection by comparing the threshold with the estimated attack. |
| [74] | Dynamic load altering attacks (D-LAA) on cyber-physical systems (CPS) | Using a real-time simulation platform. In comparing the threshold with the residual, three generators and six buses power systems to verify attack detection and reconstruction feasibility. |
| [51] | Optimal PMU placement in the grid poses a cyber threat. | The probabilistic model can be applied to assess the unobservable risk of the power grid. |

## 5. CONCLUSION

The heterogeneity and dynamic nature of the critical infrastructure resulting from the massive deployment of optical fiber communication, power line carriers, wireless communication, dedicated cables, etc., have called for additional vulnerabilities with an emerging cyber threat. We begin by reviewing how cyber-attacks affects critical infrastructures and their negative impact on grid operations. Power generation and distribution companies are the targets for cyber-attack; usually, the attacker aims to disrupt or destroy industrial control systems [82]. Secondly, we also specifically look at the application of Information technology from the perspective of the smart grid with its security implications. Finally, the authors also reviewed the various cyber-attack incidents to understand how to mitigate the attack consequences. Hence authors took a step to identify the nature of the multiple attacks associated with the critical infrastructure, considering specific vulnerabilities posed to the various aspects of the electrical infrastructure. We concluded our discussion by giving a detailed account of particular cyber-attack scenarios and their recommended mitigation solutions. Our study showed indications on identifiable cybersecurity vulnerabilities and current attack surfaces or targets in the energy generation and distribution sector as follows:

1. Load frequency control (LFC) with an unstable system frequency cannot effectively function.
2. Vulnerabilities or misconfigurations in the Master Terminal Unit (MTU) or Remote Terminal Unit (RTU) could lead to Denial of Service (DoS).
3. The inadequate energy efficiency of IoT platforms also poses a significant challenge to data transmission.

4. When an attack agent distorts sample vales (S.V.) message of IEC 6180 protocols that conveys protection and automation lead to sequential outages of power generation facilities.
5. Attackers either corrupt or obstruct power to either increase or decrease value, resulting in power outages.
6. The Power generation acquisition terminal of a new energy plant (PGATNEP), per its nature, is pre-exposed to risk of eavesdropping, Denial of service attack, forgery of control instructions, tampering of measurement data, hijacking, and interference by attackers.
7. False data injection (FDI) and Denial of service (DOS) attacks have an impact on security constraint and economic dispatch (SCED), automatic generation control (AGC), and state estimation. Attackers also exploit weak access control to corporate email services from the electrical workstations.
8. A man-in-the-middle attack abused vulnerabilities in the IEC 61850 protocol's Generic Object-Oriented Substation Event (GOOSE), resulting in cascading failures in the power grid. Demand-side resources (DSRs) responsible for providing reserve capacity and enhancing the reliability of power systems could also be affected negatively by cyber-attack.
9. Finally, we discovered that the attackers are taking advantage of the Automatic Generation Control (AGC), state estimation (S.E.), and energy management systems (EMS).

In summary, primary energy critical infrastructures areas are Load frequency control (LFC), Master Terminal Unit (MTU), Remote Terminal Unit (RTU of IEC 6180 protocols, Power generation acquisition terminal of new energy plant (PGATNEP), False measurements, Weak access control to corporate electrical workstations Generic Object-Oriented Substation Event (GOOSE) of IEC 61850 protocol, Automatic generation control (AGC), Security constraint and economic dispatch (SCED), Demand-side resources (DSRs), State estimation (S.E.) and energy management systems (EMS).

Based on our study, major cybersecurity areas that emerge are:

Denial of Service (**DoS**), the inadequate Energy efficiency of IoT platforms, eavesdropping, forgery of control instructions, tampering of measurement data, hijacking and interference injecting inaccurate measurements, false data injection (FDI), and Man in the middle.

## 6. RECOMMENDATION

Currently, a plethora of research attention has been investigating cyberattack incidents and various ways of mitigating cybersecurity-related issues in line with the critical physical infrastructure. Per our review, it is clear that the research community has been focusing on a cyber-physical aspect of the crucial infrastructure without much emphasis on exploring cybersecurity from a human behavior perspective in the energy sector. Hence the need to investigate cybersecurity on human behavior in the energy generation sector to understand how cybersecurity could impact the effective operation of critical energy infrastructure. The authors propose future work focus on how human behavior within the energy generation and distribution environment could contribute to cyber-attack in the sector and how these measures will prevent future occurrences.

# LIST OF ACRONYMS

| Acronyms | Meaning |
|---|---|
| (AMI) | Advanced Meter Infrastructure |
| (AGC) | Automatic Generation Control |
| (BESS) | Battery Energy Storage Systems |
| (IoT) | Internet of Things |
| (CPSs) | Cyber-physical |
| (CPS) | Cyber-physical systems |
| (CPPS) | Cyber-physical power systems |
| (CLCA) | Coordinated Load-Changing Attacks |
| (DSRs) | Demand-side resources |
| (DDoS) | Distributed Denial of Service |
| (DoS) | Denial-of-Service |
| (D-LAA) | Dynamic Load Altering Attacks |
| (DAF) | Dynamic Analytical Framework |
| (DGA) | Distribution Grid Automation |
| (EPS) | Electric power system |
| (EMS) | Energy Management Systems |
| (FDI) | False Data Injection |
| (FDIA) | False Data Injection Attack |
| (IIRA) | Industrial Internet Reference Architecture |
| (OSiRM) | Open systems IoT reference Model |
| (SCADA) | Supervisory Control and Data Acquisition |
| (SST) | Smart Sensor Technology |
| (ICS) | Industrial Control Systems |
| (IDS) | Intrusion Detection System |
| (ITU) | International Telecommunications Union |
| (PLC) | Programmable logic controller |
| (PLCC) | Power-line carrier communication |
| (RTU) | Remote terminal unit |
| (PMU) | Phasor measurement unit |
| (PGATNEP) | Power generation acquisition terminal of a new energy plant |
| (LFC) | Load Frequency Control |
| (GOOSE) | Generic Object-Oriented Substation Event |
| (GSM) | Global System for Mobile |
| (GPRS) | General Packet Radio Services |
| (NCCG) | Nested Column-and-Constraint Generation |
| (WSN) | Wireless Sensor Network |
| (ZigBee) | It is based on the IEEE 802.15.4 specification and is used to build networks that require a low data transfer rate, energy efficiency, and secure networking. |

# APPENDIX

Table 5. List of some of the current journals used for the analysis

| Author(s) | Journal/ Conference paper | Year of publication | Number of papers |
|---|---|---|---|
| [34], [36], [37], [63], [68] | **Conference Paper** | **2017-2020** | **5** |
| [44] | **Sustainable Energy, Grids and Networks** | **2017** | **1** |
| [48] | **J. Parallel Distrib. Comput.** | **2017** | **1** |
| [59] | **Computers and Electrical Engineering** | **2018** | **1** |
| [49][52][47] | **Electrical Power and Energy Systems** | **2018-2020** | **3** |
| [60] | **Smart Cities Cybersecurity and Privacy** | **2019** | **1** |
| [64] | **Alexandria Engineering Journal** | **2019** | **1** |
| [76][58] | **IEEE Transactions on Industrial Informatics** | **2019, 2020** | **2** |
| [54] | **Journal of Systems Architecture** | | **1** |
| [53] | **Computers & Security** | 2020 | |
| [77] | **Kaspersky ICS CERT** | **2020** | **1** |
| [62] | **IEEE Systems Journal** | 2020 | |
| [66] | **Electronic Notes in Theoretical Computer Science** | **2020** | **1** |
| [78][79] | **IEEE Access** | **2020** | **2** |
| [55][67] | **Journal of Information Security and Applications** | **2020** | **2** |
| [57] | **Electric Power Systems Research** | **2020** | **1** |
| [50][70] | **International Journal of Electrical Power and Energy Systems** | **2021** | **2** |
| [73] | **Computer Science Review** | **2021** | **1** |
| [71][51] | **Reliability Engineering and System Safety** | **2021** | **2** |
| [61] | **Applied Energy** | **2021** | **1** |
| [46] | **Automatica** | **2021** | **1** |
| [74] | **Journal of the Franklin Institute** | **2021** | **1** |
| [75] | **Blockchain: Research and Applications** | **2021** | **1** |

54

International Journal of Network Security & Its Applications (IJNSA) Vol.13, No.4, July 2021*Natl. Power Eng. Conf.*, no. October, pp. 1–4, 2018, DOI: 10.1109/NPEC.2018.8476807.

[25] A. Ghasempour, "Internet of Things in Smart Grid: Architecture, Applications, Services, Key Technologies, and Challenges," 2019, DOI: 10.3390/inventions4010022.

[26] M. Chen, J. Wan, and F. Li, "Machine-to-Machine Communications : Architectures, Standards and Applications," vol. 6, no. 2, pp. 480–497, 2012, DOI: 10.3837/tiis.2012.02.002.

[27] N. A. Hidayatullah, A. C. Kurniawan, and A. Kalam, "Power Transmission and Distribution Monitoring using Internet of Things (IoT) for Smart Grid," *IOP Conf. Ser. Mater. Sci. Eng.*, vol. 384, no. 1, 2018, doi: 10.1088/1757-899X/384/1/012039.

[28] S. Rekha, and J. Anita, "Role of smart grid in the power sector and challenges for its implementation : A review on Indian scenario," no. October 2018.

[29] S. Chakrabarty, D. W. Engels, and S. Member, "A Secure IoT Architecture for Smart Cities," 2016.

[30] A. Ramamurthy and P. Jain, "The Internet of Things in the Power Sector Opportunities in Asia and the Pacific," no. 48, 2017.

[31] Junru Lin. et al., "Monitoring Power Transmission Lines using a Wireless Sensor Network Wireless," *Commun. Mob. Comput. (John Wiley Sons, Ltd*, 2014.

[32] O. K. and Hans-RolfT., "Sensor Technology and Future Trend IEEE Transaction on Instrumentation and Measurement," *IEEE*, pp. 1497-1501.53(6) p, 2004.

[33] S. Kim, U. Kim, and J. Huh, "A Study on Improvement of Blockchain Application to Overcome Vulnerability of IoT Multiplatform Security," 2019, DOI: 10.3390/en12030402.

[34] A. Janjić, L. Velimirović, J. Ranitović, and Ž. Džunić, "Internet of Things in Power Distribution Networks – State of the Art," no. September 2017.

[35] T. Nguyen, S. Wang, M. Alhazmi, M. Nazemi, A. Estebsari, and P. Dehghanian, "Electric Power Grid Resilience to Cyber Adversaries: State of the Art," *IEEE Access*, vol. 8, pp. 87592–87608, 2020, DOI: 10.1109/ACCESS.2020.2993233.

[36] M. A. Shahid, R. Nawaz, I. M. Qureshi, and M. H. Mahmood, "Proposed Defense Topology against Cyber Attacks in Smart Grid," *4th Int. Conf. Power Gener. Syst. Renew. Energy Technol. PGSRET 2018*, no. September, pp. 1–5, 2019, DOI: 10.1109/PGSRET.2018.8685944.

[37] M. Sahabuddin, B. Dutta, and M. Hassan, "Impact of cyber-attack on isolated power system," *2016 3rd Int. Conf. Electr. Eng. Inf. Commun. Technol. iCEEiCT 2016*, pp. 8–11, 2017, DOI: 10.1109/CEEICT.2016.7873088.

[38] Kaspersky, "Cyber threats for ICS in Energy in Europe. Object of research," pp. 1–11, 2020.

[39] C. Alcaraz, P. Najera, R. Roman, and J. Lopez, "How will city infrastructure and sensors be made smart?," *White Pap.*, vol. 6, no. 11, p. 113, 2010, doi: 10.1002/047011276X.

[40] M. Eckel and T. Laffey, "Ensuring the integrity and security of network equipment is critical in the fight against cyberattacks," *Netw. Secure.*, vol. 2020, no. 9, pp. 18–19, 2020, DOI: 10.1016/S1353-4858(20)30107-0.

[41] Electric Power Research Institute (EPRI), "Contributions of Supply and Demand Resources to Required Power System Reliability Services," 2015.

[42] S. Ghosh and M. H. Ali, "Exploring Severity Ranking of Cyber-Attacks in Modern Power Grid," 2019.

[43] Z. Livingston, Sanborn, Slaughter, "Managing cyber risk in the electric power sector | Deloitte Insights," 2019.

[44] P. Eder-Neuhauser, T. Zseby, J. Fabini, and G. Vormayr, "Sustainable Energy, Grids and Networks Cyberattack models for smart grid environments," *Sustain. Energy, Grids Networks*, vol. 12, pp. 10–29, 2017, DOI: 10.1016/j.segan.2017.08.002.

[45] A. S. Bretas, N. G. Bretas, B. Carvalho, E. Baeyens, and P. P. Khargonekar, "Smart grids cyber-physical security as a malicious data attack : An innovation approach ✰," *Electr. Power Syst. Res.*, vol. 149, pp. 210–219, 2017, DOI: 10.1016/j.epsr.2017.04.018.

[46] J. Xu, B. Liu, H. Mo, and D. Dong, "Automatica Bayesian adversarial multi-node bandit for optimal smart grid protection against cyber attacks ☆," *Automatica*, vol. 128, p. 109551, 2021, DOI: 10.1016/j.automatica.2021.109551.

[47] S. Hasan, A. Dubey, G. Karsai, and X. Koutsoukos, "Electrical Power and Energy Systems A game-theoretic approach for power systems defence against dynamic," *Electr. Power Energy Syst.*, vol. 115, no. January 2019, p. 105432, 2020, DOI: 10.1016/j.ijepes.2019.105432.

[48] B. Li, R. Lu, W. Wang, and K. R. Choo, "Distributed host-based collaborative detection for false data injection attacks in smart grid cyber-physical system," *J. Parallel Distrib. Comput.*, vol. 103, pp. 32–41, 2017, DOI: 10.1016/j.jpdc.2016.12.012.
56Electronic copy available at: https://ssrn.com/abstract=3910922